\documentclass[aps,twocolumn,showpacs]{revtex4}
\usepackage{epsfig}
\usepackage{amssymb}

\begin{document}

\title{Efficiency enhancement of organic based Light Emitting Diodes using a scattering layer.}
\author{R. F. Oulton}
\author{C. S. Adjiman}
\affiliation{Centre for Process Systems Engineering \\ Department of Chemical Engineering, \\ Imperial College London.}

\author{K. Handa and S. Aramaki}
\affiliation{Mitsubishi Chemical Group \\ Science and Technology Research Centre \\ Yokkaichi, Japan.}

\begin{abstract} 

This paper presents an investigation of organic LED extraction efficiency enhancement using a low refractive index scattering layer. A scattering model is developed based on rigorous electromagnetic modelling techniques. The model accounts for proportions of scattered guided and radiation modes as well as the efficiencies with which emitted and scattered light are extracted. Constrained optimisation techniques are implemeneted for a single operation wavelength to maximise the extraction efficiency of a generic OLED device. Calculations show that a $2$ fold efficiency enhancement is achievable with a correctly engineered scattering layer. The detailed analysis of the enhancement mechanism highlights ways in which this scheme could be improved.

\end{abstract}


\maketitle

\section{Introduction}

An important consideration in the design of any optoelectronic device is the efficiency with which light can be extracted. A fundamental limitation on all light emitting materials with refractive indices greater than the collection medium is that light emitted outside the critical angle with respect to the collection medium will be trapped by total internal reflection. One successful approach that overcomes this problem is to texture the surface of devices so that trapped light can be scattered into the critical angle of extraction \cite{Sch93,Mad00,Lu01}. This approach works well for inorganic devices where the scattering surface is very close to the emission region \cite{Sch93}. In organic based LEDs the textured surface is generally separated from the emission zone by the thickness of the substrate ($\simeq 1$ mm). Although light trapped in the substrate can be extracted \cite{Shi04}, the most significant improvements require high refractive index substrates that are matched to the emission region \cite{Mad00}. Introducing wavelength size texturing closer to the active layers is generally not feasible as this would compromise the device's electrical properties.

\begin{figure}
\epsfig{file=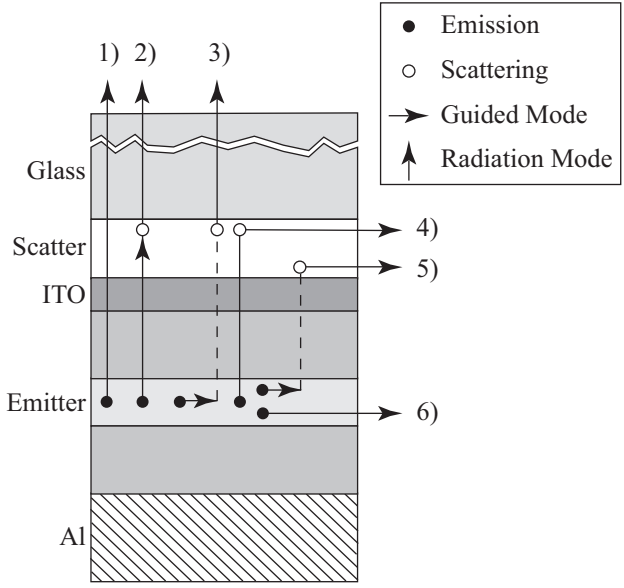}
\caption{Identification of the channels of emission for an OLED with a low refractive index scattering layer.}\label{fig1}
\end{figure}

The use of a microcavity to redistribute emission geometrically has been demonstrated in both organic \cite{Jor96} and inorganic material systems \cite{Sch96,Oul01,Ben98}. Despite their effectiveness, microcavities only provide enhancement over a limited spectral range and are significantly more elaborate in design. A more recent demonstration has shown that low refractive index hydrophobic aerogel materials can be used to successfully enhance extraction efficiency in organic based LEDs \cite{Tsu01}. In this approach, trapped light from an emitting region ($n=1.8$) is encouraged to couple evanescently through the higher index Indium Tin Oxide (ITO) anode ($n_{ITO} \approx 1.9$) to a thick aerogel layer ($n<1.1$). Angular resolved intensity enhancements near a factor of $1.5$ were reported; likely to be less in terms of actual power enhancement. A theoretical study of volumetric light scattering in such low refractive index substrates shows that significant improvements can result in combining these two enahcnement mechanisms \cite{Shi04}. Here a similar approach is examined which combines the successful texturing / scattering approache with the use of low refractive index layers. A scattering layer, consisting of high refractive index dielectric spheres embedded in a low refractive index material, is placed beside the anode of an organic device as depicted in Fig.~\ref{fig1}. This is possible because of the surface flatness and uniformity of the scattering layer, developed by Mitsubishi Chemicals Corporation, that does not interfere with the electrical properties of the device. The distance between the emission region and this layer is comparable to the wavelength of light allowing efficient scattering of low order guided modes that constitute a large proportion of trapped light. Scattered light is effectively redistributed within the lower refractive index medium from where it can be extracted with greater efficiency.
 
Consider the channels through which emission is extracted or trapped within the device. Light originates via spontaneous emission from the emission region in either radiation modes which can escape the device with efficiency, $\eta_c(\lambda)$, or guided modes that are trapped, $(1 - \eta_c(\lambda))$. Proportions of radiation, $\gamma_R(\lambda)$, and guided, $\gamma_G(\lambda)$, modes are then redistributed into the same set of radiation and guided modes inside the scattering layer. Scattered radiation modes are extracted from the scattering layer with efficiency $\eta_s(\lambda)$.  Figure~\ref{fig1} identifies the $6$ emission channels whose descriptions and resulting efficiencies are given below.

\begin{enumerate}

\item Unscattered radiation modes: $\eta_c(\lambda)(1-\gamma_R(\lambda))$.

\item Radiation modes scattered into radiation modes: $\eta_c(\lambda)(1-\gamma_R(\lambda))\eta_s(\lambda)$.

\item Guided modes scattered into radiation modes: $(1-\eta_c(\lambda))\gamma_G(\lambda)\eta_s(\lambda)$.

\item Radiation modes scattered into guided modes: $\eta_c(\lambda)\gamma_R(\lambda)(1-\eta_s(\lambda))$.

\item Guided modes scattered into guided modes: $(1-\eta_c(\lambda))\gamma_G(\lambda)(1-\eta_s(\lambda))$.

\item Unscattered guided modes: $(1-\eta_c(\lambda))(1-\gamma_G(\lambda))$.

\end{enumerate}

Let $\eta_c^{(0)}(\lambda)$ be the extraction efficiency when the scattering strength is zero. The first order efficiency, $\eta^{(1)}_c(\lambda)$, is given by the sum of those emission channels that result in the extraction of radiation modes. When $\eta^{(1)}_c(\lambda)$ is greater than $\eta^{(0)}_{c}(\lambda)$, the scattering layer provides an improvement in efficiency. Eqn.~(\ref{eqn1}) expresses this inequality in a slightly different form.
\begin{eqnarray}\label{eqn1} \frac{\gamma_G(\lambda)}{\gamma_R(\lambda)} > \frac{\eta^{(0)}_c(\lambda)(1-\eta_s(\lambda))}{\eta_s(\lambda)(1-\eta^{(0)}_c(\lambda))} \end{eqnarray}

Given that the extraction efficiency from a dielectric medium into air is approximated by $1 - \cos \theta_c$ where $\theta_c$ is the critical angle with respect to air. This assumes both uniform emission and scattered light distributions. Eqn.~(\ref{eqn1}) can be approximated by,
\begin{eqnarray}\label{eqn2} \frac{\gamma_G(\lambda)}{\gamma_R(\lambda)} > \frac{\left(\frac{n_c }{ \sqrt{n_c^2 - 1}} - 1\right)}{\left(\frac{n_s }{ \sqrt{n_s^2 - 1}} - 1\right)} \end{eqnarray}

where $n_c$ and $n_s$ are the refractive indices of the emitting and scattering layers respectively. The efficiency enhancement of the device is summarised by Eqns.~(\ref{eqn1}) and (\ref{eqn2}): The LHS must be maximised while the RHS must be minimised. Minimisation of the RHS of Eqns.~(\ref{eqn1}) and (\ref{eqn2}) is achieved by reducing $n_s$ with respect to $n_c$. Choice of a material with as low a refractive index as possible is clearly required. The LHS of Eqns.~(\ref{eqn1}) and (\ref{eqn2}) is maximised by ensuring that guided modes are scattered more effectively than radiation modes.  

By substituting $g \gamma_G(\lambda) = g\gamma(\lambda) = \gamma_R(\lambda)$ into Eqn.~\ref{eqn2}, the first order extraction efficiency enhancement, $\eta^{(1)}(\lambda) / \eta^{(0)}_{c}(\lambda)$, is given by,
\begin{eqnarray}\label{eqn2a} \frac{\eta^{(1)}(\lambda)}{\eta^{(0)}_{c}(\lambda)} = f \approx 1+ \gamma(\lambda) \left[ \frac{\cos \theta_c(1-\cos \theta_s)}{(1- \cos \theta_c)} - g \cos \theta_s \right] \end{eqnarray}

Here, the distribution of emitted and scattered light is assumed to be uniform and $\theta_s$ and $\theta_c$ are the critical angles for coupling into air from the scattering and emitting layers respectively. As an example, consider the redistribution of scattered light from the emitting region with index $n_c = 1.8$ to the scattering region with $n_s = 1.1$; this would give a first order efficiency enhancement of $f \approx 1 + \gamma(\lambda) (2.88 - 0.42 g )$. When $\gamma(\lambda) = 0.5$ and guided modes are scattered equally as effectively as radiation modes ($g = 1$), the enhancement factor $f = 2.34$. Note that this would also be augmented by multiple scattering. In the limiting case of $\gamma(\lambda) \mapsto 1$, the efficiency $\eta_c^{(1)}(\lambda) \mapsto \eta_s(\lambda)$ and the enhancement factor, $f_{lim} \mapsto 3.46$. 

The prospect of such large enhancement factors to the optical extraction efficiency of OLEDs is very appealing. The calculations in this paper show that the limiting value of the enhancement factor for a more realistic model is $f_{lim} \mapsto 2$. Furthermore, similar enhancements are achievable across the visible spectrum. With the current fabrication techniques employed by Mitsubishi Chemical Corporation, $75$ \% of this value can be attained. This paper investigates an optimised scattering layer OLED design. The four model components introduced in Eqns.~(\ref{eqn1}) and~(\ref{eqn2}) are evaluated using rigorous electromagnetic methods. The focus of these calculations centres on the treatment of guided mode scattering as this is the crucial component of the optimisation scheme.

\section{Organic LED design.}

In the following study, a simplified generic OLED design is investgated. The aim is to make rigorous calculations of the parameters discussed in the preceeding section in order to obtain better estimates of the merits of the efficinecy enhancement strategy. Table~\ref{tab1} gives the refractive indices and layer thicknesses for a typical OLED device. The basic structure of the device has been optimised to enhance the extraction of radiation modes to a first order scattering approximation at a wavelength of $\lambda = 450$ nm. The augmentation of the efficiency due to the scattering layer will be investigated by controlling three parameters, specified in Table~\ref{tab1}.  The parameters are the scattering layer thickness $d_s$, refractive index $n_s$ and the hole conduction layer thickness, $d_h$. During the optimisation, the scattering coefficient, $\alpha_s$, is kept constant. Currently, Mitsubishi have successfully engineered scattering layers with $\alpha_s d_s = 0.33$ so $\alpha_s = 0.33/d_s$. The constrained optimisation problem is formally stated in Eqn.~\ref{eqn3}.
\begin{eqnarray}\label{eqn3}  \max_{d_s,\,n_s,\,d_h} & \eta^{(1)}_{c}(d_s,\,n_s,\,d_h) & \nonumber \\ s.t. & 200 \leq  d_s  \leq 800 & \nonumber  \\ & 1.1 \leq  \Re\{ n_s\}  \leq 1.4 & \nonumber  \\ & 50 \leq  d_h  \leq 200 &  \end{eqnarray}

\begin{table}[t]
\begin{tabular}{|l||c|c|} \hline
Layer Description & Thickness (nm) & Ref. Index @ $450$ nm \\ \hline \hline
Air  				&  $\infty$	&  1		\\ \hline 
Glass  				&  Incoherent	&  1.54 \cite{Pal}	\\ \hline
\textbf{Scattering layer}$^\dagger$	&  $d_s=311$&  $n_s = 1.1 - \alpha_s\lambda / 2 \pi^*$ \\ \hline
Anode ITO			&  150 		&  1.978 - 1.73x10$^{-2}$i\cite{web1}	\\ \hline
\textbf{Hole Conductor}  	&  $d_h=79$	&  1.8$^*$	\\ \hline
Emission (Alq3)		&  30		&  1.8$^*$	\\ \hline
Electron Conductor  		&  50		&  1.8$^*$	\\ \hline
Cathode Al  			&  100		&  0.62 - 5.47i \cite{Pal}	\\ \hline
Air				&  $\infty$	&  1 		\\ \hline
\end{tabular}
\caption{Table specifying the OLED design under investigation. Since the glass substrate is $0.7$ mm thick, it is modelled as an incoherent layer. Here the refractive indices are given for $450$ nm. $^*$ indicates a refractive index value that has been either determined or estimated experimentally.}\label{tab1}
\end{table}

\begin{figure}
\epsfig{file=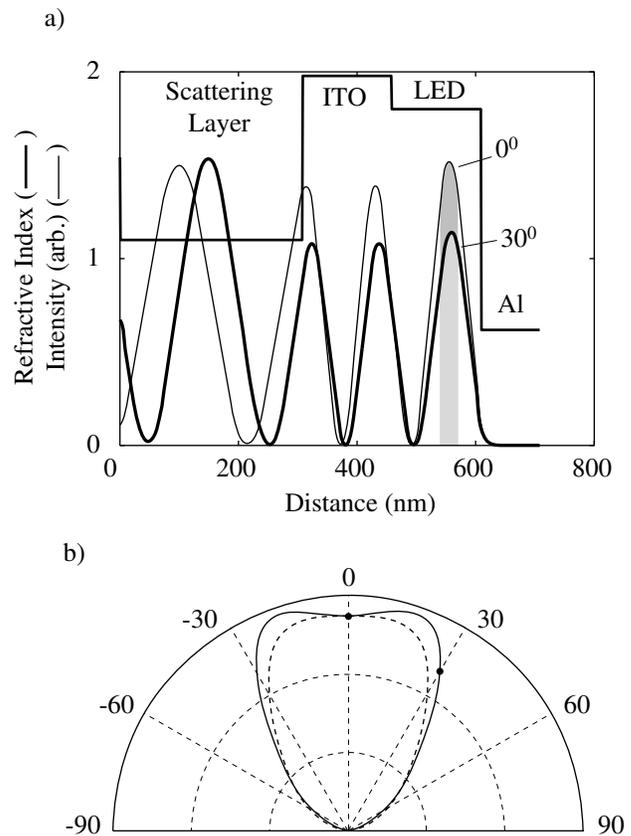}
\caption{(a) Radiation fields within the OLED: Solid line shows the radiation field at normal incidence and the broken line shows the Transverse Electric (TE) polarisation at $30^\circ$ with respect to air. Notice in particular the field overlap at the shaded emission region. (b) The radial plot shows the strength of the electric field intensity in the emission region as a function of emission angle. Points corresponding to the field plots in Fig.~\ref{fig2} (a) are indicated by circular markers.}\label{fig2}
\end{figure}

Here, the objective function is the efficiency from a first order scattering enhancement. Although multiple scattering has been considered later in the report, introducing it in the objective function seriously increases the computation time of local solutions. A global solution was determined by successive optimisation searches using distributed starting points spanning the parameter space. A total of $125$ device configurations were considered resulting in three local maximum solutions. The global maximum solution is shown in Table~\ref{tab1} and is used throughout the paper.

The operation of optical devices in the spontaneous emission regime involves spectral, spatial and angular parameterisations. For example, Figure~\ref{fig2} (a) shows the electric field intensity at normal incidence and $30^\circ$ for the Transverse Electric (TE) polarisation within the device at a wavelength of $\lambda = 450$ nm corresponding to blue light. The strength of the optical field and its overlap with the emission zone of the device, highlighted by the shaded area, determine the light extraction from the device. This is plotted as a function of angle in Fig.~\ref{fig2} (b). This level of detail makes the analysis very complicated.

In the present study, the four efficiency parameters are calculated by integrating over solid angles leaving only the wavelength as the independent variable. This simplifies the problem greatly, but, it is important to remain mindful of the internal angular and spatial variations. In addition, the polarisation is eliminated from the study by averaging over dipole orientations. For details of the techniques and models incorporated here, the reader is referred to Ref. \onlinecite{Ben98a}. 

\section{Evaluation of model components.}

\subsection{Evaluation of the underlying extraction efficiency, $\eta^{(0)}_c$}

The extraction efficiency, $\eta^{(0)}_c$, of the OLED device is calculated by integrating angular emission results, such as those in Fig.~\ref{fig2} (b), over solid angle for $\alpha_s = 0$. 

Consider, the impact of the optimum solution on $\eta^{(0)}_c$: Notice that interference effects due to the large contrast between the ITO and scattering layer will result in weak cavity effects that enhance $\eta_c$. This is observed in Fig.~\ref{fig2} (b), which shows that emission at $17.3^\circ$ is preferred in the optimised design. Lobed emission like this is a result of the weak cavity being detuned from the preset wavelength of $450$ nm. This is preferential as there is a larger density of off-axis emission states \cite{Oul01}. The tuning wavelength of the weak cavity is given approximately by $\lambda_0 = 450 / \cos \theta_{max}$ where $\theta_{max}$ is the angle of maximum emission. Here, $\lambda_0 = 456.3$ nm.

Evidently, interference effects are important in the determination of this model component as the thickness of the scattering layer is a multiple of $\lambda_0/4$. Here, $3\lambda_0/4n_s \approx 311.1$ nm $\approx d_s$.
 
\subsection{Evaluation of the scattered light extraction efficiency, $\eta_s$}

The extraction efficiency of scattered light, $\eta_s$, can be calculated provided the scattered light distribution within the layer is known. Here, the distribution of scattered light is assumed to be uniform. This is a good approximation provided scatterers are much smaller than the wavelength of light. This can be shown for small spherical dielectric scatterers using Mie theory \cite{Book}. 

Following a scattering event, light is collected from the low refractive index layer. Since subsequent scattering events are considered separately in Section~\ref{sec2g}, the complex part of the scattering layer refractive index is dropped for the calculation of $\eta_s$. Under these condiions, $\eta_s$ represents the upper limit on extraction from the scattering layer. As expected from the discussion in the introduction, the optimum choice of refractive index in the scattering layer corresponds to the lower bound.

\subsection{Evaluating the proportion of scattered radiation mode power, $\gamma_R$}\label{sec2e}

$\gamma_R$ is a measure of the strength of radiation mode scattering within the device. It is calculated by considering the difference between a device with ($\alpha_s \neq 0$) and without ($\alpha_s = 0$) scattering.  Where there is no scattering, the extraction efficiency is $\eta^{(0)}_c$. With scattering, the zero order extraction efficiency is reduced to $\eta_{c(\alpha)} = (1-\gamma_R)\eta^{(0)}_c$. Therefore, $\gamma_R$ is given by Eqn.~(\ref{eqn4}).
\begin{eqnarray}\label{eqn4} \gamma_R =\frac{ \eta^{(0)}_{c}- \eta_{c(\alpha)}}{\eta^{(0)}_c}  \end{eqnarray}

\subsection{Evaluating the proportion of scattered guided mode power, $\gamma_G$.}\label{sec2f}

\begin{figure}
\epsfig{file=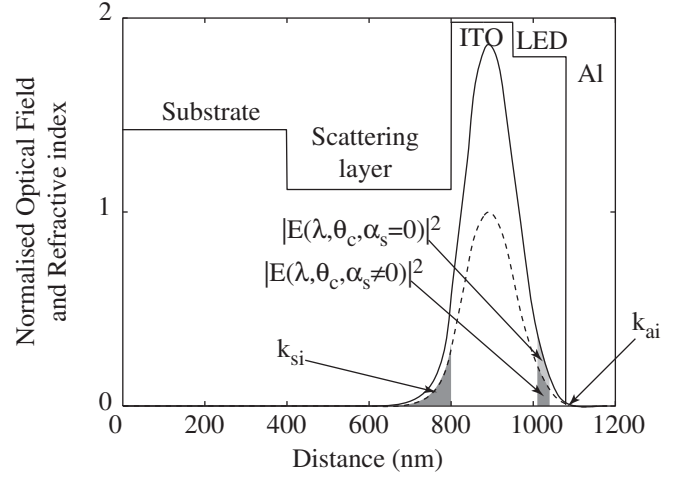}
\caption{Illustration of important features of the $\gamma_G(\lambda)$ calculation. Here, the absorption, emission and scattering of the TE$_1$ mode is shown, where the Solid Line corresponds to $|E(\lambda,\alpha_s = 0)|^2$ and the broken line to $|E(\lambda,\alpha_s \neq 0)|^2$. $k_{si}(\lambda)$ allows the calculation of mode attenuation due to scattering. Absorption, $k_{ai}(\lambda)$, is strongest at the metal surface.}\label{fig3}
\end{figure}

The proportion of scattered guided modes, $\gamma_G$ is evaluated by examining the internal fields of the device. Despite this, the calculation approach is similar to the evaluation of $\gamma_R$. The optical field intensity, $|E(\lambda,\theta_c,\alpha_s = 0)|^2$, within the emission region in the absence of scattering ($\alpha_s = 0$) can be directly compared wtih the field intensity, $|E(\alpha_s,\theta_c \neq 0)|^2$, for $\alpha_s \neq 0$ to give a value for $\gamma_{G}$. Note that this does not take into account the lateral extent of the device, although this will be considered shortly. Fig.~\ref{fig3} illustrates the relevant components in these guided mode scattering calculations for the TE$_1$ mode corresponding to a particular internal angle, $\theta_c$. $\gamma_G$ is evaluated using the expression in Eqn~\ref{eqn5} by integrating over the internal emission solid angle, $\Omega_c$.
\begin{eqnarray}\label{eqn5} \gamma_G =  \frac{\int_{\Omega_c}|E(\alpha_s=0)|^2 - |E(\alpha_s \neq 0)|^2d\Omega_c}{\int_{\Omega_c}|E(\alpha_s=0)|^2d\Omega_c}   \end{eqnarray}

\begin{figure}
\centering
\epsfig{file=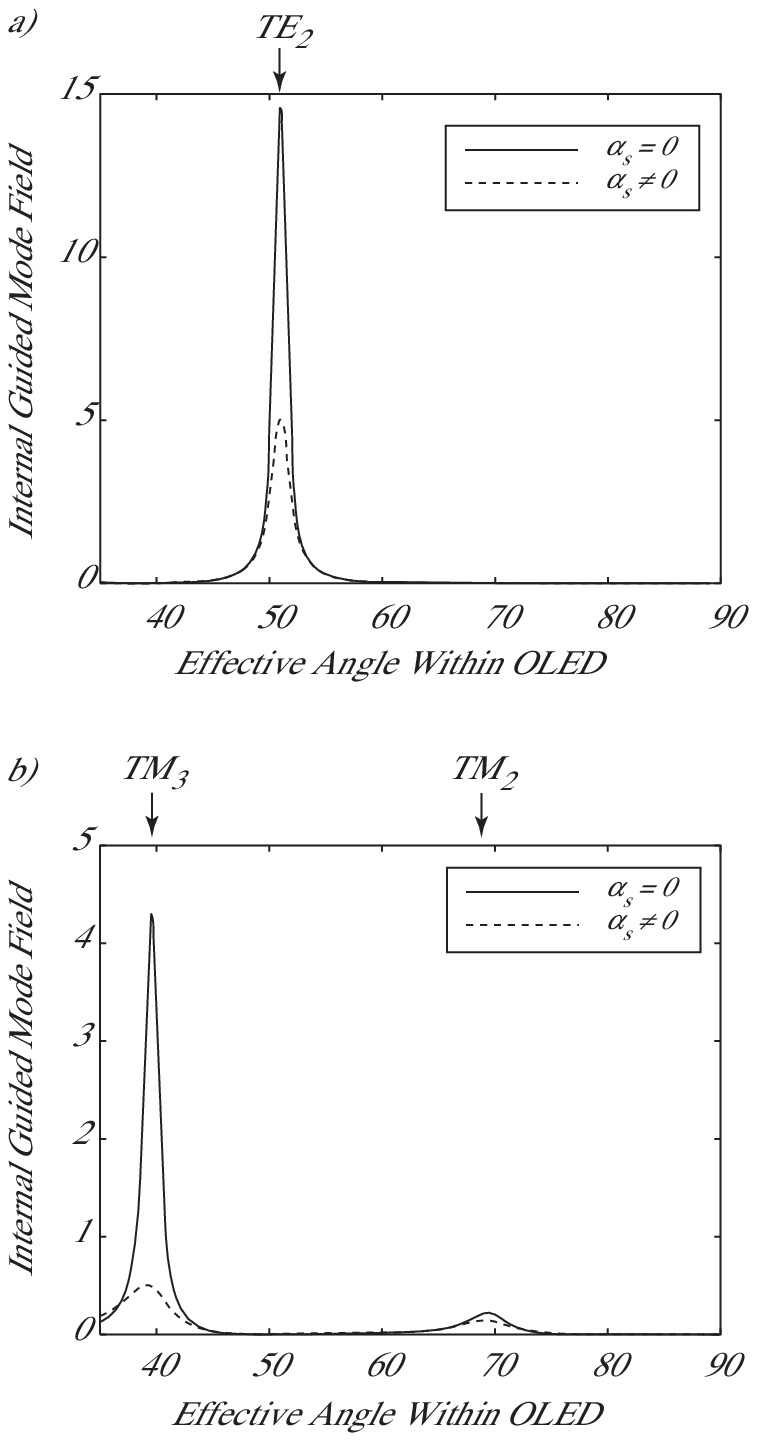}
\caption{Examination of the scattering of guided modes in an effective plane wave angle representation for (a) TE and (b) TM polarisations. $|E(\lambda,\alpha_s\neq 0)|^2$ are represented by solid lines and $|E(\lambda,\alpha_s=0)|^2$ by broken lines showing the change in guided mode coupling with the OLED due to scattering at $\lambda = 450$ nm.}\label{fig4}
\end{figure}

Figure~\ref{fig4} plots $|E(\lambda,\theta_c,\alpha_s = 0)|^2$ and $|E(\alpha_s,\theta_c \neq 0)|^2$ highlighting the effect of the scattering layer on the optical field within the device against the internal angle, $\theta_c$. Here, $90^\circ$ corresponds to propagation in the plane of the device with respect to the light emitting region ($n = 1.8$). TE emission is shown in Fig.~\ref{fig4} (a) while TM emission is shown in Fig.~\ref{fig4} (b). The arrows indicating the peaks correspond to the solutions of guided modes considered later; the effective mode angles (angles at peaks in optical field) correspond to those given in Table~\ref{tab2}. The amount of scattering can be gauged by the difference between the broken lines ($\alpha_s = 0$) and the solid lines ($\alpha_s \neq 0$). In this case, when integrated over all trapped emission, $\gamma_{G} \approx 35$ \%. 

A mode $i$ propagating within the device has a propagation constant, $\beta_i$. $\beta_i$ is in general complex and is calculated using the Argument Principle Method \cite{Ane94,Ane99}. The complex propagation constant of a mode describes the phase velocity of propagation in the plane of the device and the mode attenuation as it propagates, given by the real and imaginary parts respectively. The imaginary components of the propagation constant $k_{s\, i} + k_{a\, i}$ for the $i^{th}$ guided mode will indicate mode attenuation due to scattering and absorption respectively. $k_{s\, i}$ can be evaluated by subtracting modal propagation constants calculated with and without scattering, $\alpha_s$.

Table~\ref{tab2} shows the complex propagation constants, $\beta_i(\alpha_s)$, for the optimised OLED device specified in Tab~\ref{tab1} for the scattering layer with $\alpha_s = 0$ and $\alpha_s \neq 0$. Notice that the internal mode angles, $\theta_{c}$, relative to the emission region are shown and correspond directly to the indicated modes in Fig.~\ref{fig4}.

\begin{table*}
\begin{tabular}{|c||c|c|c|c|c|} \hline
 Mode  &  $\beta_i(\alpha_s = 0$) $[nm^{-1}]$  &  $\beta_i(\alpha_s \neq 0$) $[nm^{-1}]$  & $k_{s\, i}$  &   $\theta_c$&  $L$  \\ \hline \hline
TM$_{3}$ & $1.550 \times 10^{-2} - 5.398 \times 10^{-5}i$ & $1.550 \times 10^{-2} - 1.590 \times 10^{-3}i$ & $ - 1.54 \times 10^{-3}i$ & $38.07^\circ$ & $0.7$ $\mu$m \\ \hline
TE$_{2}$ & $1.953 \times 10^{-2} - 1.601 \times 10^{-4}i$ & $1.953 \times 10^{-2} - 2.592 \times 10^{-4}i$ & $ - 9.90 \times 10^{-5}i$ & $51.99^\circ$ & $10.1$ $\mu$m \\ \hline
TM$_{2}$& $2.321 \times 10^{-2} - 3.205 \times 10^{-4}i$ & $2.321 \times 10^{-2} - 4.134 \times 10^{-4}i$ & $ - 9.29 \times 10^{-5}i$ & $67.44^\circ$ & $10.8$ $\mu$m \\ \hline
TE$_{1}$ & $2.535 \times 10^{-2} - 1.986 \times 10^{-4}i$ & $2.535 \times 10^{-2} - 2.188 \times 10^{-4}i$ & $ - 2.03 \times 10^{-5}i$ & $>90^\circ$ & $49.3$ $\mu$m \\ \hline
TM$_{1}$& $2.678 \times 10^{-2} - 3.427 \times 10^{-4}i$ & $2.678 \times 10^{-2} - 3.471 \times 10^{-4}i$ & $ - 4.40 \times 10^{-6}i$ & $>90^\circ$ & $227.3$ $\mu$m \\ \hline \hline
\end{tabular}
\caption{Propagation constants, $\beta_i$ for the most important modes of an OLED. The two sets of results (with and without scattering) allow calculation of $k_{s\, i}$. In addition, the internal mode angles, $\theta_c$, with respect to the emitting region are also given.}\label{tab2}
\end{table*}

\begin{figure}
\centering
\epsfig{file=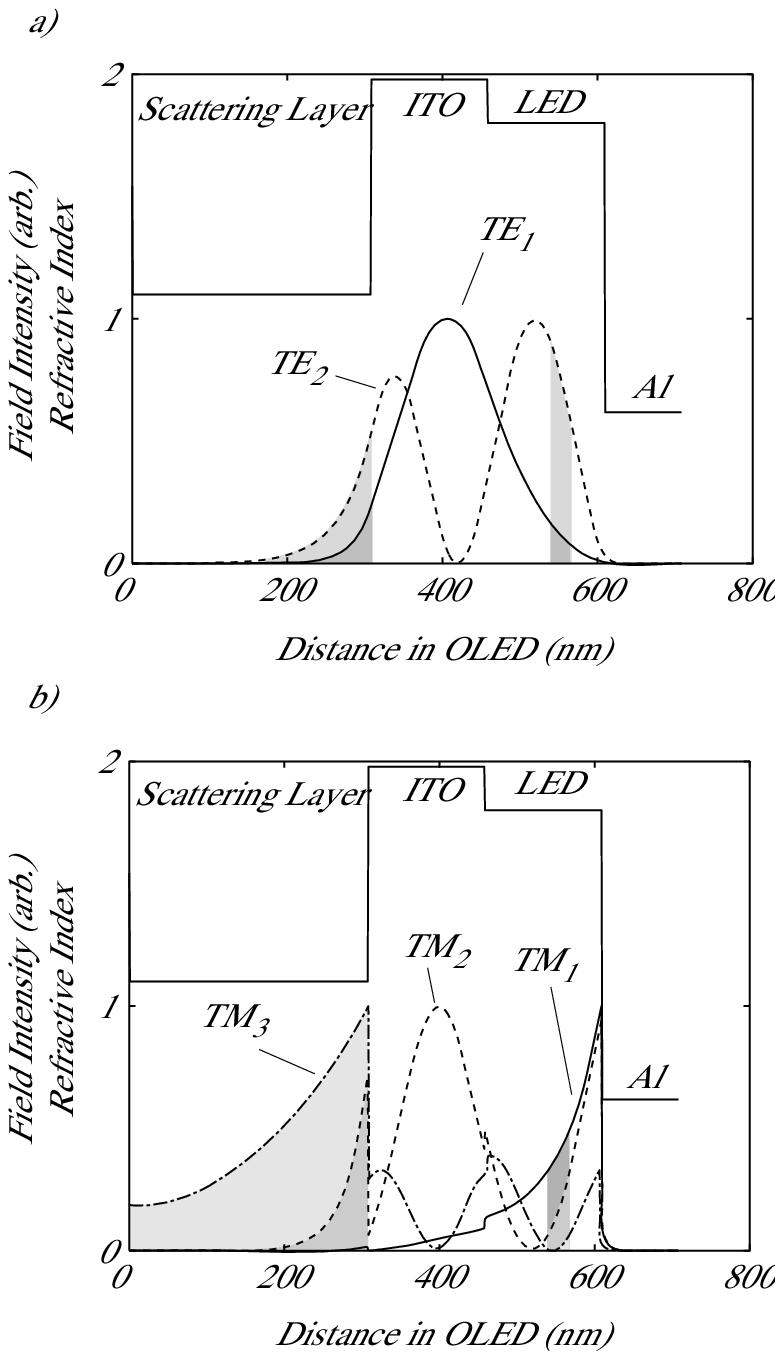}
\caption{Field distributions of standing modes of the OLED corresponding to peaks in field intensity in Fig.~\ref{fig4} for (a) TE and (b) TM polarisations. A detailed parameterisation of these modes including effective mode angles, extinction coefficients and absorptions are shown in Tab.~\ref{tab2}.}\label{fig5}
\end{figure}

The complex eigenvalues in Tab.~\ref{tab2} correspond to a guided mode eigenvector within the OLED. The electric field distribution for the modes are shown in Fig.~\ref{fig5}. Note that the eigenvalues of the mode functions calculated here correspond to the peaks in emission distribution in Figs.~\ref{fig4} (a) and~\ref{fig4} (b). The mode functions indicate both the degree of absorption and the scattering in the different regions of the device. Firstly, the two polarisations are quite distinct: TE modes propagate away from the metal interface in the dielectric with the highest refractive index, whereas TM modes propagate along the surface of the metal. Absorption within the ITO region has an equally profound effect on both polarisations contributing significantly to $k_{s\, i}$. Fig.~\ref{fig5} highlights two specific regions of interest: The electric field intensity overlap with the emission and scattering regions. It is evident that the overlap of TM modes with the emitting region is very poor in contrast to the TE modes explaining the different electric field intensity $y$-axis scales in Figs.~\ref{fig4} (a) and~\ref{fig4} (b).  However, the overlap of the field with the scattering region is also critical as it indicating the strength of guided mode scattering. The TE$_1$ and TM$_1$ have poor overlaps with both the scattering layer and the emitting region. In addition any emissive coupling to these mode is purely evanescent since there critical angles are greater than $90^\circ$ with respect to the emission region. Only the higher order TE$_2$, TM$_2$ and TM$_3$ modes are required to describe the mode scattering and absorption.

Close examination of the values in Tab.~\ref{tab2} shows that absorption is stronger than the scattering. This is reflected by the field calculations which shows that $\gamma_G \approx 35$ \%. For laterally small devices some of the trapped light may not be scattered or absorbed. Here, $k_{s\, i}$ and $k_{a\, i}$, can be used to estimate the distance a guided mode travels before being fully scattered. The total mode attenuation length, $L$, shown in Tab.~\ref{tab2}, is defined as the distance a mode travels before being scattered to $e^{-1}$ of its original intensity by scattering alone (i.e. calculated using $k_{s\,i}$). Typically, a device with a lateral size of the order of $10\, \mu$m is large enough to maximise guided mode scattering at $\lambda = 450$ nm.

\subsection{Calculating multiple scattering.}\label{sec2g}

The device efficiency due to first order scattering can be expressed as,
\begin{eqnarray}\label{eqn6} \eta^{(1)}_c = M_1\eta^{(0)}_c + C_1\end{eqnarray}

where $M_1= 1 - \gamma_R+ \eta_s(\gamma_R - \gamma_G)$ and $C_1 = \gamma_G\eta_s$. The efficiency for higher order scattering can similarly be expressed as,
\begin{eqnarray}\label{eqn7} \eta^{(n)}_c = M_n\eta^{(n-1)}_c + C_n \end{eqnarray}

$M_n$ and $C_n$ are given by expressions similar to $M_1$ and $M_2$, however, use the new model components $\gamma^{(2)}_G$ and $\gamma^{(2)}_R$ to quantify the proportions of light that are scattered a second time and subsequent time. These parameters have been calculated using similar techniques to those described in Sections~\ref{sec2e} and~\ref{sec2f}. In the case of the optimisation calculation, where $\alpha_s d_s = 0.33$, multiple scattering contributes $5$ \% of the efficiency.

\section{Results.}

\begin{figure}
\centering
\epsfig{file=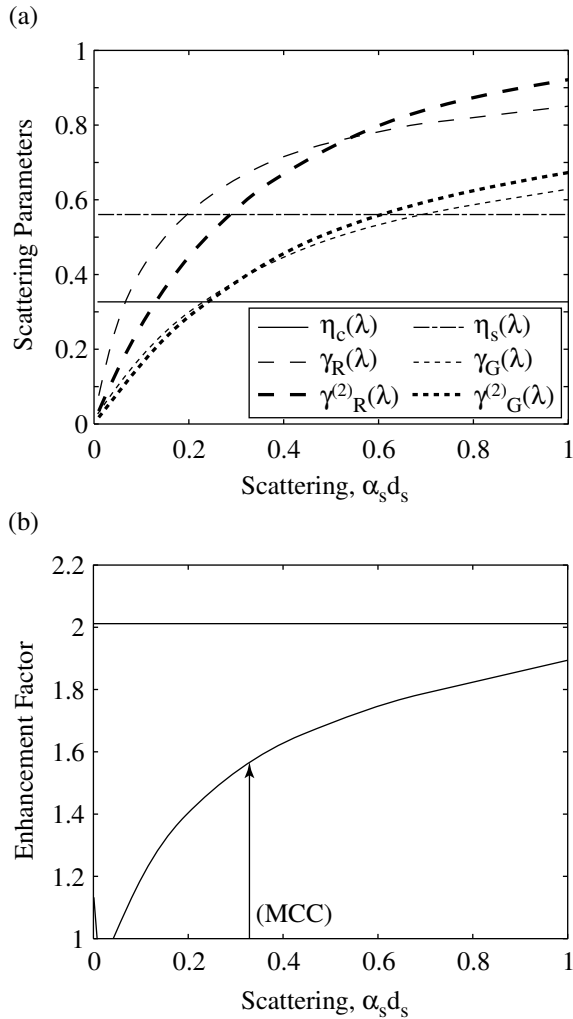}
\caption{(a) Variation of model components $\eta_c(\lambda)$, $\eta_s(\lambda)$, $\gamma_R(\lambda)$ and $\gamma_G(\lambda)$ as a function of scattering strength. (b) Variation of enhancement factor as a function of scattering strength. The trend is indicative of linear scattering theory (see Eqn.~(\ref{eqn8})). The resultant efficiencies of the device with and without the scattering layer are also shown for comparison.}\label{fig6}
\end{figure}

So far the internal components of the calculations have been examined in detail. Now consider the behaviour of the model components as a function of the scattering strength. The strength of scattering ultimately governs the degree of mixing between the radiation and guided modes of the device. As the scattering is increased, there will also be an increase in enhancement factor of the underlying device. This behaviour is seen in Fig.~\ref{fig6}.

Fig.~\ref{fig6} (a) shows the variation of the model components as a function of the scattering strength, $\alpha_s d_s$ whose range is extended to $0 \leq \alpha_s d_s \leq 1$. The extraction efficiencies, $\eta_c$ and $\eta_s$ are independent of the scattering strength. The proportions of scattered guided and radiation modes increase monotonically with the scattering strength and differ in magnitude by a factor $g \approx 1.25$. Clearly, radiation modes are scattered more strongly that guided modes. This is a critical observation since to acquire the greatest efficiency enhancement, this ratio must be minimised.

Fig.~\ref{fig6} (b) plots of the efficiency enhancement factor, $f(\alpha_s d_s)$, for the case of multiple scattering. Here, the enhancement factor has been compared to the efficiency of a device without a scattering layer, $\eta_{c0} = 27.8$ \%. The enhancement factor $f(\alpha_s d_s)$ is given by the following expression related to Eqn.~\ref{eqn2a}.
\begin{eqnarray}\label{eqn8} f(\alpha_s d_s) = \frac{\eta^{(0)}_c}{\eta_{c0}} + \left(   \frac{\eta_s}{\eta_{c0}} - (1 + g(1-\eta_s))\frac{\eta^{(0)}_{c}}{\eta_{c0}} \right)\gamma + \mathcal{O}(\gamma^2) ...  \end{eqnarray}

In the absence of absorption, there is an enhancement of about $20$ \% due to the presence of the low index scattering layer, {\it c.f.} the first term in Eqn.~(\ref{eqn8}). This is comparable to enhancements obtained by Tsutsui et al in their experiments with aerogel layers \cite{Tsu01}. As the scattering strength is increased, there is an initial reduction in efficiency, mainly due second order scattering loss in the low refractive index region. Above $\alpha_sd_s = 0.05$, however, $f(\alpha_s d_s)>1$ and increases monotonically. For the current design an enhancement that is $95$ \% of the limiting value of $\eta_s/\eta_{c0}$ is attained. This is still almost a two-fold enhancement in the efficiency. With the current engineering capabilities of Mitsubishi Chemicals Corporation, nearly $75$ \% of the upper limit can be achieved, corresponding to a $60$ \% enhancement.

The first order scattering provides the largest contribution to the enhancement factor. Indeed the first order scattering coefficient in Eqn.~(\ref{eqn8}) encapsulates the limitations of this enhancement approach. The primary limitations, in order of importance, are the extraction efficiency from the low refractive index medium, $\eta_s$, the strength of guided mode scattering, $\gamma$, and the ratio of radiation to guided mode scattering, $g$. $\eta_s$ must be maximised, requiring a scattering layer with as low a refractive index as possible. Here, the optimum value of $n_s=1.1$ is at the lower bound of what can be achieved in the fabrication process. Alternative materials such as aerogels, which have refractive indices as low as 1.01 \cite{Tsu01}, would show even larger enhancements. However, the choice of material must be compatible with the formation of dielectric spheres to provide the required scattering strength, $\gamma$. Maximising $\gamma$ allows the device to attain a larger fraction of the limiting enhancement factor. Finally, the ratio of guided to radiation modes, $g$ must be minimised.

In Sec.~\ref{sec2f} the calculation of the guided mode scattering component, $\gamma_G$, is detailed. The reader is therefore referred to this section for an in depth appraisal of the factors that effect the value of $g$. One of the most significant factors in the guided mode scattering strength is the strong absorption in the high refractive index anode region (ITO) to which modes are confined. At $\alpha_s d_s = 0.33$, guided mode absorption is approximately $3/2$ times larger than scattering. In contrast, scattering is approximately $4$ times larger than the absorption for the radiation modes. Low absorption near the active components of the device is clearly crucial for minimising $g$ and maximising the efficiency enhancement.

The principles of operation presented here suggest enhancements of up to a factor of $2$ could be achieved with a carefully designed device incorporating a scattering layer. Although this is comparable to microcavity enhancements \cite{Jor96}, here, the enhancement is achievable across the visible spectrum. The design of an enhancement layer for a broad spectral range would be limited by the scattering strength drop-off at red wavelengths and the difficulty associated with maximising coherent reflections. Despite this, calculation of the device structure investigated here using scattering data from Mitsubishi  at green and red wavelengths show $f(\lambda = 550) = 1.5$ and $f(\lambda = 630) = 1.3$. In addition, the device design could be optimised for overall broad spectral performance.

\section{Conclusions.}

A perturbative model was developed for the description of low refractive index scattering layers that enhance the extraction efficiency of light from organic LEDs. Components of the model were calculated using rigorous electromagnetic techniques. The scattering model was used to optimise an OLED design incorporating a scattering layer. The calculations show that a two-fold enhancement in the extraction efficiency is attainable.

Three parameters were highlighted as crucial to the enhancement mechanism. Most importantly a low refractive index medium that supports high refractive index scattering particles is necessary to set the limiting enhancement factor and the scattering strength required to attain it. Finally, the ratio of radiation to guided modes must be minimised. Optimisation of the first two parameters is difficult as they depend on complex fabrication techniques. In contrast, careful device design could allow the ratio of radiation guided mode scattering to be reduced. 

\acknowledgements

This work is supported by Mitsubishi Chemicals Corporation. Thanks also to Dr. P. Stavrinou for useful discussions.

\break

\end{document}